\documentclass[preprint,showpacs,preprintnumbers,amsmath,amssymb]{revtex4}

\usepackage{graphicx}
\usepackage{dcolumn}
\usepackage{bm}

\begin{document}

\def\be{\begin{equation}}
\def\ee{\end{equation}}
\def\bea{\begin{eqnarray}}
\def\eea{\end{eqnarray}}

\preprint{ } \vskip .5in
\title{Stabilization of the Extra Dimensions in Brane Gas Cosmology with Bulk Flux }

\author{Jin Young Kim\footnote{Electronic address:
jykim@kunsan.ac.kr}}
\address{Department of Physics, Kunsan National University,
Kunsan 573-701, Korea, and \\Asia Pacific Center for Theoretical
Physics, Pohang 790-784, Korea}
\date{\today}

\begin{abstract}

We consider the anisotropic evolution of spatial dimensions and
the stabilization of internal dimensions in the framework of brane
gas cosmology. We observe that the bulk RR field can give an
effective potential which prevents the internal subvolume from
collapsing. For a combination of $(D-3)$-brane gas wrapping the
extra dimensions and 4-form RR flux in the unwrapped dimensions,
it is possible that the wrapped subvolume has an oscillating
solution around the minimum of the effective potential while the
unwrapped subvolume expands monotonically. The flux gives a
logarithmic bounce to the effective potential of the internal
dimensions.

\end{abstract}



\maketitle

\section{Introduction}

There has been considerable effort during the past decade in the
area of higher dimensional gravity where the four-dimensional
gravity can be recovered as an effective theory. In the theory of
general relativity, the dimensionality of the universe is an
assumption. It is not derived dynamically from a fundamental
theory. However, unifying gravity with other forces of nature
strongly suggests that there may be more than three spatial
dimensions. One plausible explanation can be given by string
theory, which is higher dimensional. A mechanism to generate three
spatial dimensions was proposed in Refs. \cite{bv} and \cite{tv}.
This is a mechanism to generate dynamically the spatial
dimensionality of spacetime and to explain the problem of initial
singularity. The key ingredient of this model is based on the
symmetry of string theory, called T-duality. With this symmetry,
spacetime has a topology of a nine-dimensional torus, and its
dynamics is driven by a gas of fundamental strings. Since strings
have (1+1)-dimensional world volumes, they can intersect in 2(1+1)
spacetime dimensions or less from the topological point of view.
The winding modes can annihilate the anti-winding modes if they
meet. Once the windings are removed from the dimensions, these
dimensions can expand into the large directions that we observe
today. Thus, with a gas of strings, three spatial dimensions can
become large.

With the realization of D-branes in string theory \cite{polch},
there had been attempts to study string cosmology with a gas of
D-branes including strings \cite{magrio,psl}.  The mechanism of
Brandenberger and Vafa (BV) in Refs. \cite{bv,tv} was reconsidered
in this new framework of string theory. Authors of Ref. \cite{abe}
argued that when branes are included, strings still dominate the
evolution of the universe at late times. This model of brane gas
cosmology was studied extensively
\cite{bgextended,cbbc,kayarador,egjk1,jykim,modulstab,campos,Cheungwatsonbran}
(see also \cite{bW0510022} for comprehensive reviews). In the
picture of brane gas cosmology, the universe starts from a hot,
dense gas of D-branes in thermal equilibrium. The winding modes of
branes prevent the spatial dimensions from growing. Branes with
opposite winding numbers can annihilate if their world volumes
intersect. Thus, a hierarchy of scales can be achieved between
wrapped and unwrapped dimensions. Though the BV mechanism can
trigger the onset of cosmological evolution based on string
theory, this does not guarantee the answer to the question why
there are three large spatial dimensions while others remain
small. There have been attempts to yield the anisotropic expansion
based on certain wrappings of the brane gases \cite{egjk1,jykim}.

It is well known that the stability of internal dimensions cannot
be achieved with string or brane gases of purely winding modes.
Stabilizing the extra dimensions is an important issue in string
or brane gas cosmology
\cite{modulstab,campos,Cheungwatsonbran,patil1,patil2}. For a
particular configuration of brane gas in eleven dimensional
supergravity, Campos \cite{campos} has shown the importance of
fluxes in stabilizing the internal dimensions through numerical
computations of the evolution equations for wrapped and unwrapped
dimensions. It is shown that all internal dimensions, except the
dilaton, are dynamically stabilized in the presence of two-form
flux with a gas of long strings \cite{Cheungwatsonbran}. In
certain string theories \cite{patil2}, special string states can
stabilize the extra dimensions for fixed dilaton. Massless string
modes at self-dual radius play a crucial role for the radion
stabilization.

In this work we will focus on the anisotropic evolution of spatial
dimensions and the stabilization of internal dimensions based on a
certain configuration of branes and bulk background field. We
consider the possible stabilization of internal dimensions with
bulk Ramond-Ramond (RR) field. We observe that the bulk RR field
can give an effective potential which prevents the internal volume
from collapsing. The flux gives a logarithmic bounce to the
effective potential of the internal volume.

In Sec. II, we consider the brane universe in the frame of
Einstein gravity. We present a formalism for anisotropic expansion
of brane universe and show that it is impossible to stabilize the
internal volume with brane gases only. In Sec. III, we set up a
supergravity formalism with brane gas and bulk RR field. For the
anisotropic expansion, we consider the global rotational symmetry
is broken down to $SO(p) \times SO(D-p)$ by $(D-p)$-brane gas. We
decouple the equations of motion for each subvolumes of $SO(p)$
and $SO(D-p)$ and argue the possible stabilization. In Sec. IV, we
conclude and discuss.

\section{Anisotropic expansion of brane universe}

In the previous work of us \cite{jykim}, we have shown that the
extra dimensions should be wrapped with three- or
higher-dimensional branes for the observed three large spatial
dimensions. However, it is not guaranteed that the compact
dimensions can be stabilized with these branes. We start, after
the thermal equilibrium of brane gas is broken, from the case when
three dimensions are unwrapped and $D-3$ dimensions are wrapped by
gas of branes whose dimensions are less than or equal to $D-3$. We
assume that each type of brane gases makes a comparable
contribution to the energy momentum density. Then, we can study
the evolution of a universe based on the following form of metric
 \be
 g_{MN} = {\rm diag} ( -1, a^2 \delta_{ij} , b^2 \delta_{mn} ),
 \label{asymmetric}
 \ee
where $a^2$ is the scale factor of the $3$-dimensional space and
$b$ is the scale factor of the internal $(D-3)$-dimensional
subspace.

The evolution of the universe with the above setting is equivalent
to the asymmetric inflation in the brane world scenario
\cite{adkmr}. We will consider a universe which can be described
by the effective action
 \be S = \int d^{D+1}x \sqrt{g} \Big [ \frac{R}{2\kappa^2} -
V_{D+1}(g_{\mu\nu})\Big ] , \label{effact}
 \ee
 where $\kappa^2 = 1/ M^{D-1}_*$,
$M_*$ being the $(D+1)$-dimensional unification scale, and
$V_{D+1}(g_{\mu\nu})$ is the effective potential for the metric
generated by the breakdown of the Poincare invariance. Considering
the spatial section to be a $D$-dimensional torus $T^D$, we write
the metric as
 \be
ds^2 = -n^2 dt^2 + \sum_k a^2_k(t) (dx_k )^2 . \label{metric}
 \ee
Technically, we retain the lapse $n$ because we work on the action
to reduce the action to a one-dimensional effective action. When
we vary the reduced action with respect to the lapse, we obtain
the Hamiltonian constraint. Later, we will set $n=1$. Also we use
the metric (\ref{metric}) instead of metric (\ref{asymmetric}) to
produce the correct combinational factors when the partial
derivatives are taken over different scale factors. After all the
derivatives are evaluated, we will take the identities $a_1 =
\cdots = a_3 = a$ and $a_4 = \cdots = a_D = b$.

The effective potential $V_{D+1}$ arising from the brane tensions
can be written as
 \be
 \sqrt{g} V_{D+1}(g_{\mu\nu}) = n \Big [ \Lambda_D \prod_{k=1}^D a_k +
  \sum_{\{p\}} T_{\{p\}} \prod_{\{p\}} a_k +  \prod_{k=1}^D a_k U (a_k )
  \Big ] ,  \label{potential}
 \ee
 where $\Lambda_D$ is the tension of a $D$-dimensional space-filling
 brane which can be considered as a bulk cosmological constant,
$T_{\{p\}}$ is the tension of $p$-brane averaged over space, and
$U (a_k ) $ parameterizes all contributions to the potential other
than branes. For example, Fluxes, curvatures, and brane-brane
interactions can give non-zero terms to $U(a_k )$. We will show
that $U (a_k )$ plays an important role for the internal
dimensions from collapsing to zero. In the summation over the
brane tension, the notation $\{p\}$ refers to the fact that there
may be branes with the same $p$ oriented along different
directions. For example, 2-brane can wrap (12) cycle, (23) cycle
et cetera.

Now, we evaluate the Ricci scalar in Eq. (\ref{metric}) and reduce
the problem to the motion of a particle in $D$ dimensions. The
one-dimensional effective action for the scale factors $a_k$ is
 \bea
 S_{\rm eff}
 &&= -{\cal V}_D \int dt ~n\prod_{k=1}^D a_k
  \Big[
\frac{1}{2\kappa^2} \Big \{ (\sum_{j=1}^D \frac{ \dot a_j}{
na_j})^2 - \sum^D_{j=1} (\frac{\dot a_j}{na_j})^2 \Big \}
\nonumber \\
&&~~~~~~~~~~~~~~~~~~~~~~~~~~ + \Lambda_D +\frac{1}{\prod_{n=1}^D
a_n } \sum_{\{p\}} T_{\{p\}} {\prod_{\{p\}} a_k } + U (a_k ) \Big]
, \label{actred}
 \eea
where ${\cal V}_D = \int d^Dx$ is the coordinate spatial volume,
which contains the branes, and is dimensionless. In these units,
the scale factors $a_k$ are dimension-full quantities, with
dimension $M^{-1}_*$. If we vary this action with respect to $n$
and $a_k$, we obtain the following set of equations of motion
after some straightforward algebra and setting $n=1$ at the end,
 \be
 (\sum_{j=1}^D \frac{\dot a_j}{a_j})^2 -
 \sum_{j=1}^D (\frac{\dot a_j}{a_j})^2
 = 2\kappa^2 \{ \Lambda_D + U (a_k ) \}
 + \frac{2\kappa^2}{\prod_{k=1}^D a_k} \sum_{\{p\}} T_{\{p\}}
 \prod_{\{p\}} a_k , \label{eomcon}
 \ee
 \bea
  \frac{\ddot a_n}{a_n} + \sum_{l\ne n}
\frac{\dot a_n \dot a_l}{a_n a_l}
 &=& \frac{ 2 \kappa^2}{D-1} \{ \Lambda_D + U (a_k ) \} +
\frac{\kappa^2}{(D-1)\prod_{k=1}^D a_k} \sum_{\{p\}} T_{\{p\}}
\prod_{\{p\}} a_k
\nonumber \\
&+& \frac{\kappa^2}{(D-1)\prod_{k=1}^D a_k} \sum_{l=1}^D
\sum_{\{p\}} T_{\{p\}} a_l \frac{\partial}{\partial
a_l} \prod_{\{p\}} a_k  \nonumber \\
 &-& \frac{\kappa^2}{\prod_{k=1}^D a_k} \sum_{\{p\}} T_{\{p\}} a_n
\frac{\partial}{\partial a_n} \prod_{\{p\}} a_k
 - \kappa^2 a_n \frac{\partial U }{\partial a_n}
   + \frac{\kappa^2}{D-1} \sum_{k=1}^D
   a_k \frac{\partial U }{\partial a_k} . \label{eoms}
 \eea
We emphasize here that the partial derivatives are meant to be
taken over different scale factors, even if some of the scale
factors may coincide on the background. Note that the derivative
operator $a_l \frac{\partial}{\partial a_l}$ behaves like a
counting operator acting on $\prod_{\{p\}} a_k$. If $\prod_{\{p\}}
a_k$ depends on $a_l$, the operator produces $1$; otherwise, it
gives $0$. Hence, the operator $\sum_l a_l
\frac{\partial}{\partial a_l}$ simply counts the dimension of the
brane.

Now, we consider the special case when the global rotational
symmetry is broken down to $SO(3) \times SO(D-3)$ by effective
$(D-3)$-branes whose tension we denote $T_{D-3}$. We ignore the
stabilizing potential $U(a_k )$ for simplicity. In this case, the
equations of motion in Eqs. (\ref{eomcon}) and (\ref{eoms}) are
simplified to, taking units in which $2\kappa^2=1$,
 \be
6(\frac{\dot a}{a})^2 + (D-3)(D-4) (\frac{\dot b}{b})^2 + 6(D-3)
\frac{\dot a \dot b}{ab} = \Lambda_D +  \frac{T_{D-3}}{a^3},
 \label{constraintequation}
 \ee
 \be
 \frac{\ddot a}{a} + 2(\frac{\dot a}{a})^2 + (D-3)
\frac{\dot a \dot b}{ab}
 = \frac{\Lambda_D}{D-1}  +
\frac{D-2}{2(D-1)} \frac{T_{D-3}}{a^{3}},  \label{simpleoma}
 \ee
 \be
 \frac{\ddot b}{b} + (D-4)(\frac{\dot b}{b})^2 + 3 \frac{\dot
a \dot b}{ab} = \frac{\Lambda_D}{D-1} - \frac{1}{2(D-1)}
\frac{T_{D-3}} { a^{3}} . \label{simpleomb}
 \ee
The largest difference between the sizes $a$ and $b$ can be
induced if $a$ grows faster than $b$, as dictated by Eqs.
(\ref{simpleoma}) and (\ref{simpleomb}). The key parameters for
controlling the relative rates of growth of $a$ and $b$ are their
accelerations, not their velocities. To ensure that $a$ exceeds
$b$ by many orders of magnitude, the sources in Eqs.
(\ref{simpleoma}) and (\ref{simpleomb}) must produce the slow-roll
conditions for $b$, making its acceleration small or negative
while keeping the acceleration of $a$ positive.

If the volumes of wrapped and unwrapped subspaces are defined as
 \be
 \zeta \equiv a^3 ,~~~~ \xi \equiv b^{D-3},
 \ee
the second derivative terms in Eqs. (\ref{simpleoma}) and
(\ref{simpleomb}) can be written as
 \bea
 \frac{\ddot \zeta}{\zeta} + \frac{\dot \zeta \dot \xi}{\zeta
\xi} &=& 3 \Big \{ \frac{\Lambda_D}{D-1} + \frac{D-2}{2(D-1)}
\frac{T_{D-3}}{\zeta} \Big \} ,
\label{neweomzeta}   \\
 \frac{\ddot \xi}{\xi}  + \frac{\dot \zeta \dot \xi}{\zeta \xi}
 &=& (D-3) \Big \{ \frac{ \Lambda_D}{D-1} -
\frac{1}{2(D-1)} \frac{T_{D-3}}{\zeta} \Big \} . \label{neweomxi}
 \eea
The above equations of motion reduce to a simple form by removing
the coupled first-derivative terms:
 \be
\frac{\ddot \xi}{\xi} -  \frac{D-6}{D-1} \Lambda_D = \frac{\ddot
\zeta}{\zeta} - \frac{4D-9}{2(D-1)} \frac{T_{D-3}}{\zeta}.
\label{xizetadecoupled}
 \ee
Since the left-hand side of Eq. (\ref{xizetadecoupled}) is a
function of $\xi$ while the right-hand side is a
 function of $\zeta$, each side should be the same constant. Taking the
 constant as the parameter $E$, we have the following set of decoupled
 equations for $\zeta$ and $\xi$:
 \bea
 \frac{\ddot \zeta}{\zeta} &=& E + \frac{4D-9}{2(D-1)}
 \frac{T_{D-3}}{\zeta} ,     \label{zetadd}   \\
 \frac{\ddot \xi}{\xi} &=& E + \frac{D-6}{D-1}
 \Lambda_D.     \label{xidd}
 \eea
Note that the constant $E$ is not an independent parameter. For
the given configuration of branes, it is related to brane tension
$T_{D-3}$ and cosmological constant $\Lambda_{D}$. The constant
$E$ can be found by putting the solutions of (\ref{zetadd}) and
(\ref{xidd}) into the original equation (\ref{neweomzeta}) and
(\ref{neweomxi}).

To make the unwrapped subvolume $\zeta$ expand forever, we require
the constant $E$ to be a positive definite. Then, the solution for
the subspace $\zeta$ is
 \be
 \zeta = ( \zeta_0 +C_1 )  e^{\gamma t} - C_1,   \label{zetaexpon}
 \ee
 where $ \zeta_0$ is the volume of unwrapped subspace at $t=0$, $\gamma = \sqrt{E}$, and
 $ C_1 = \frac{4D-9}{2(D-1)} \frac{T_{D-3}}{E} $.
For given $E$, three types of solution for $\xi$ are possible
depending on the bulk cosmological constant $\Lambda_D$. First,
for $\Lambda_D>0$, we have
 \be
\xi = \xi_0 e^{\gamma_1 t}, ~~\gamma_1 = \sqrt{E + \frac{D-6}{D-1}
\Lambda_D }.     \label{fasterxi}
 \ee
 Since $\gamma_1 > \gamma$, $\xi$ expands faster than $\zeta$. This
means that the wrapped dimensions expand faster than the unwrapped
dimensions. To determine the constant $E$, we put the solutions
(\ref{zetaexpon}) and (\ref{fasterxi}) into (\ref{neweomzeta})
 \be
 \gamma^2 (1 + \frac {C_1} {\zeta} )
  + \gamma \gamma_1 ( 1 + \frac {C_1} {\zeta} )
 = 3 \Big \{ \frac {\Lambda_{D}} {D-1}
 + \frac {D-2} {2(D-1)} \frac {T_{D-3}} {\zeta} \Big \}.
 \ee
This gives the following relation between the constant $E$ and
other parameters in the unit of $2 \kappa^2 = 1$,
 \bea
 \gamma^2 + \gamma \gamma_1  &=& 3 \frac { \Lambda_D } {D-1}
 , \label{paracase1-1}   \\
  ( \gamma^2 + \gamma \gamma_1 ) C_1 &=& \frac {3(D-2)} {2(D-1)} T_{D-3} .
   \label{paracase1-2}
 \eea
There is no positive value of $E$ which can suffice both of the
above equations. This means that the solutions of the types
(\ref{zetaexpon}) and (\ref{fasterxi}) are not possible.

Second, for $ \Lambda_D = - |\Lambda_D |< 0$ and $E -
\frac{D-6}{D-1} |\Lambda_{D}| > 0$, the possible solution for
$\xi$ is of the form
 \be
 \xi = \xi_0 e^{\gamma_2 t} , ~~
 \gamma_2 = \sqrt{E - \frac{D-6}{D-1}
 |\Lambda_{D}| }.  \label{slowerxi}
 \ee
In this case, $\xi$ expands slower than $\zeta$.  Repeating the
same procedure as in the previous case, to check whether this
solution can be realized, we have
 \bea
 \gamma^2 + \gamma \gamma_2  &=& - 3 \frac { |\Lambda_D |} {D-1}
 , \label{paracase2-1}    \\
  ( \gamma^2 + \gamma \gamma_2 ) C_1 &=& \frac {3(D-2)} {2(D-1)} T_{D-3} .
   \label{paracase2-2}
 \eea
 Since $E>0$ is a strict condition for the expansion of the unwrapped dimensions,
 it is obvious that the first condition cannot be satisfied.

Finally, for $ \Lambda_D = - |\Lambda_D |< 0$ and $E -
\frac{D-6}{D-1} |\Lambda_{D}| < 0$, we have
 \be
  \xi = \xi_0 \cos( \omega t ),~~
 \omega = \sqrt{- E + \frac{D-6}{D-1}
 |\Lambda_{D}| }.  \label{oscxi}
 \ee
In this case, the wrapped internal subvolume $\xi$ oscillates
while the unwrapped volume $\zeta$ expands monotonically. The
solution seems quite similar to the one found by Patil and
Brandenberger \cite{patil1} in string gas cosmology in five
dimensional spacetime without dilaton. They found that an
oscillating type of solution, where the radius of extra dimension
oscillates at the self-dual radius, is one of the possible
trajectories. To check whether there is any constant $E$ which
gives this possibility, we substitute (\ref{zetaexpon}) and
(\ref{oscxi}) into (\ref{neweomzeta})
 \be
 \gamma^2 (1 + \frac {C_1} {\zeta} )
  + \omega \gamma( 1 + \frac {C_1} {\zeta} ) \tan \omega t
 = 3 \Big \{ \frac {- |\Lambda_{D}|} {D-1}
 + \frac {D-2} {2(D-1)} \frac {T_{D-3}} {\zeta} \Big \}.
 \ee
It obvious that there is no nontrivial solution. We conclude, from
the results for the three cases, that it is not possible to
stabilize the internal dimensions with only branes.

The above analysis with effective $(D-3)$-branes is a special case
of the dynamics described by Eq. (\ref{eoms}). Other distributions
of brane sources can give different possibilities. However, to
have three large observed dimensions, we cannot neglect the
potential $U( a_k )$ which represents the potential other than
branes. Considering the motion of two subvolumes $\zeta$ and $\xi$
as motion of particles in one-dimension effective potentials, the
crucial point is that there is no bouncing potential for small
$\xi$ for the third case in which we are interested. So the size
of the internal subvolume is not bounded from below. This means
that the internal subvolume can shrink to zero and can have even
negative value, violating the assumption of the late-time brane
gas approximation.

\section{Brane gas cosmology with bulk RR field}

Having established the formalism and shown that it is not possible
to stabilize the internal volume with only brane gas, let us
consider a model where the potential $U(a_k )$ comes from bulk
flux. The model we will consider is type II string theory
compactified on $T^9$. Also we consider the late stage of the BV
scenario where the radii and curvature scales are grown larger
than the ten-dimensional Planck length. Since winding numbers are
topologically conserved, only branes with opposite orientations
can cause unwinding. This unwinding interactions are very short
ranged since this happens when the world volumes of branes and
antibranes physically intersect as in BV scenario \cite{bv}. When
the radii of the unwound directions have grown enough, the brane
and antibrane densities will be diluted so that they cannot find
each other to intersect. Then we can neglect the brane-antibrane
annihilation since this term will be turned off as the transverse
dimensions expand \cite{kayarador,egjk1,jykim}. Supergravity is a
good approximation with the growing radii and falling temperature.
In brane gas cosmology it is assumed that brane mode fluctuations
are small and only winding modes dominate the cosmological
evolution. So we consider the wrapped D-branes, which do not
interact with each other, for the matter part of the Einstein
equation. This corresponds to a classical approximation of the
brane dynamics. Assuming that the brane gases are homogeneous, we
can take the average of the contributions of all kinds of branes
to the energy-momentum tensor.

In the point of string theory, the gravitational interaction is
described by the coupled system of the metric and dilaton. Dilaton
plays an important role in the large-small symmetry of string
theory called T-duality. The initial big-bang singularity can be
resolved in terms of T-duality \cite{bv,tv}. We start from the
following bulk effective action of type II string theory
 \be
S_{\rm b} =\frac{1}{2\kappa^2} \int d^{D+1}x \sqrt{-g} \Bigl[
e^{-2 \phi} \{ R + 4 (\nabla \phi)^2 \} - \frac{1}{2 l!} F_l^2 -
V(\phi) \Bigr], \label{bulkea}
 \ee
 where $\phi$ is the dilaton field, $F_{l}$ is an RR $(l)$-form
 field strength and $V(\phi)$ is the dilaton potential.
The dilaton potential is a product of nonperturbative effects. For
simplicity, we consider only flat potential which equivalently is
a cosmological constant.

The matter contribution of a single brane to the action is
represented by the Dirac-Born-Infeld (DBI) action of $p$-brane
 \be
S_{\rm p} = - T_p \int d^{p+1} \xi e^{-\phi} \sqrt{ - {\rm det} (
{\hat g}_{\mu\nu} + {\hat B}_{\mu\nu} + 2 \pi \alpha^\prime
{F}_{\mu\nu} ) } , \label{pbraneea}
 \ee
where ${\hat g}_{\mu\nu}$ is the induced metric to the brane
 \be
 {\hat g}_{\mu\nu} = g_{MN} \frac{\partial x^M}{ \partial \xi^\mu}
\frac{\partial x^N}{ \partial \xi^\nu}.
 \ee
Here $M,N$ are the indices of bulk spacetime and $\mu,\nu$ are
those of brane. ${\hat B}_{\mu\nu}$ is the induced antisymmetric
tensor field and ${F}_{\mu\nu}$ is the field strength tensor of
gauge fields $A_\mu$ living on the brane. The fluctuations of the
brane coordinates and other fields within the brane are negligible
when the temperatures are low enough. So we neglect ${\hat
B}_{\mu\nu}$ and ${F}_{\mu\nu}$ terms below. The total action can
be written as
 \be
 S_{\rm tot} = S_{\rm b} + \sum_{\{ p \}} S_{\{\rm p \}} .
 \ee

Since we are interested in the late time cosmology, we have to
work in the Einstein frame defined by $ g_{MN}^{\rm S} =
e^{\frac{4}{D-1} \phi} g_{MN}^{\rm E}$. The formalism and physical
picture for the investigation of metric perturbations have been
developed in the Einstein frame where one can use the standard
simple Einstein equations. Since the weak coupling limit (i.e. $
g_s = e^{\phi} << 1$) is one of the basic assumptions of
string/brane gas cosmology, we ignore the running of dilaton
assuming that dilaton can be stabilized. When the dilaton is not
running, the two frames are equivalent since the contribution of
constant dilaton can be absorbed in the redefinition of brane
tension and RR field. Then, the action can be written as
 \be
S_{\rm b}^E =\frac{1}{2\kappa^2} \int d^{D+1}x \sqrt{-g^E} \Bigl[
R  - \frac{1}{2l!} F_l^2 - \Lambda_D
 \Bigr], \label{bulkeaEin}
 \ee
 \be
S_{\rm p} = - T_p \int d^{p+1} \xi
\sqrt{ - {\rm det}  {\hat
g}_{\mu\nu}^E } . \label{pbraneeaEin}
 \ee
 We drop the superscript $E$ from now on.

In the point of bulk theory, the energy momentum tensor of a
single D$p$-brane has a delta function singularity at the position
of the brane along the transverse directions
 \be
 \sqrt{-g} T_p^{MN} = - T_p
 \int d^{p+1} \xi  \sqrt{ - {\hat g} }
 {\hat g}^{\mu\nu} \partial_\mu x^M \partial_\nu x^N \delta
 [x - x(\xi)] .
 \ee
In cosmological setting, it seems natural to take a gas of such
branes in a continuum approximation and this smooths the
singularity by integrating over the transverse dimensions
\cite{kayarador}.

 The induced metric on the $p$-brane follows
from its embedding in the bulk. We choose the static gauge
$\xi^\mu = x^\mu$ ($\mu = 0,1, \dots,p$) and suppose the simple
embedding $x^i = x^i (t)$ for $i = p+1, \dots, D$. The induced
metric on the brane is,
 \be
 {\hat g}_{\mu\nu} = {\rm diag} \{ -(1- v^2 ), a_1^2 , a_2^2 , \dots,
a_p^2  \} ,
 \ee
 where $v^2 = \sum_{i = p+1}^{D} {\dot x}^i {\dot x}_i \ge 0$.
Note that the brane time, defined by $d \tau = dt \sqrt{1 - v^2}$,
is not the same as the bulk time if the brane is moving. For the
RR field we take the ansatz
 \be
 F_{N_1 , \cdots , N_l} = n \sqrt{l} \epsilon_{N_1 , \cdots , N_{l-1} }
  \nabla_0 A_{\{l\}} (t) , \label{Fansatz}
 \ee
 where $N_1 , \cdots , N_{l-1} \ne 0 $. Then the Bianchi identity
 \be
  \nabla_{[N} F_{N_1 , \cdots , N_l ]} = 0 , \label{Bianchi}
 \ee
is automatically satisfied since $A_l (t) $ is a function of $t$
only.

If we consider the static brane, i.e., consider the brane does not
move in the transverse direction ($v=0$), the brane time is the
same as the bulk time. Then the one-dimensional effective action
for the scale factors can be written as
 \bea
 S_{\rm eff} = - {\cal V}_D \int dt ~n\prod_{k=1}^D a_k
\Bigl[ \frac{1}{2\kappa^2} && \Bigl\{(\sum_{j=1}^D \frac{ \dot
a_j}{ na_j})^2 - \sum^D_{j=1} (\frac{\dot a_j}{na_j})^2 \Bigr\}
\nonumber \\
 &&+ \Lambda_{D} +\frac{1}{\prod_{n=1}^D a_n } \sum_{\{p\}}
T_{\{p\}} {\prod_{\{p\}} a_k } +
 \frac{1}{2} \sum_{ \{ l \} }
 \frac{{\dot A}_{ \{ l \} }^2}{ \prod_{j=1}^{l-1} a_j^2 }
 \Bigr].  \label{eacd}
 \eea
We take variation of this action with respect to $n$, $a_k$,
 and ${\dot A}_{ \{ l \} }$. After some straightforward algebra setting
$n=1$ and $2\kappa^2 = 1$ at the end, we obtain the following set
of equations of motion:
 \be
 (\sum_{j=1}^D \frac{\dot
a_j}{a_j})^2 - \sum_{j=1}^D (\frac{\dot a_j}{a_j})^2
 = \Lambda_{D} + \frac{1 }{\prod_{k=1}^D a_k}
\sum_{\{p\}} T_{\{p\}} \prod_{\{p\}} a_k
  + \frac{1}{2} \sum_{ \{ l \} }
 \frac{ {\dot A}_{ \{ l \} }^2 }{ \prod_{j=1}^{l-1} a_j^2 },
 \label{eomncd}
 \ee
 \bea
 && \frac{\ddot a_n}{a_n} + \sum_{l\ne n} \frac{\dot a_n \dot
 a_l}{a_n a_l} = \frac{1}{2(D-1)} \Bigl\{ 2 \Lambda_{D}
 +\frac{1}{\prod_{k=1}^D a_k} \sum_{\{p\}} T_{\{p\}} \prod_{\{p\}} a_k
 + \sum_{ \{ l \} }
 \frac{ {\dot A}_{ \{ l \} }^2 }{ \prod_{j=1}^{l-1} a_j^2 }
   \Bigr\} \nonumber \\
 &&
+ \frac{1}{2(D-1)} \Bigl\{ \frac{1} {\prod_{k=1}^D a_k }
\sum_{m=1}^D \sum_{\{p\}} T_{\{p\}} a_m \frac{\partial}{\partial
a_m} \prod_{\{p\}} a_k
 + \frac{1}{2} \sum_{m=1}^D a_m \frac{\partial}{\partial a_m}
 \sum_{ \{ l \} }
 \frac{ {\dot A}_{ \{ l \} }^2 }{ \prod_{j=1}^{l-1} a_j^2 }
 \Bigr\}  \nonumber  \\
&& - \frac{1}{2} \Bigl\{ \frac{1} {\prod_{k=1}^D a_k }
\sum_{\{p\}} T_{\{p\}} a_n \frac{\partial}{\partial a_n}
\prod_{\{p\}} a_k
 + \frac{1}{2} a_n \frac{\partial}{\partial a_n}
 \sum_{ \{ l \} }
 \frac{ {\dot A}_{ \{ l \} }^2 }{ \prod_{j=1}^{l-1} a_j^2 }
  \Bigr\} ,
  \label{eomancd}
 \eea
  \be
  \frac{d}{dt} \Bigl\{ \prod_{k=1}^D a_k
  \frac{  {\dot A}_{ \{ l \} } }{ \prod_{j=1}^{l-1} a_j^2 }
  \Bigr\} = 0 . ~~~~~~~~~~~~~~~~~~~~~~~~~~~~~~~~~~~~~~~~~~~~ \label{adoteqcd}
 \ee
Note that, comparing Eqs. (\ref{eomncd}) and (\ref{eomancd}) with
Eqs. (\ref{eomcon}) and (\ref{eoms}), the potential $U$ comes from
the ${\dot A}_{\{ l\} }$ term originated from bulk RR field.

The solution of (\ref{adoteqcd}) is given by
 \be
 {\dot A}_{ \{ l \} } =
  \frac{Q_{ \{ l \} } \prod_{j=1}^{l-1} a_j^2 }
  { \prod_{k=1}^D a_k } , \label{soladot}
 \ee
with an integration constant $Q_{ \{ l \} }$. Substituting
(\ref{soladot}), the equations of motion
(\ref{eomncd})-(\ref{eomancd}) can be written as
 \be
  (\sum_{j=1}^D \frac{\dot
a_j}{a_j})^2 - \sum_{j=1}^D (\frac{\dot a_j}{a_j})^2   =
\Lambda_{D} + \frac{1 }{\prod_{k=1}^D a_k} \sum_{\{p\}} T_{\{p\}}
\prod_{\{p\}} a_k
  + \frac{1}{2} \sum_{ \{ l \} } Q_{ \{ l \} }^2
   \frac{ \prod_{j=1}^{l-1} a_j^2 }
 {\prod_{k=1}^D a_k^2} ,
 \label{eomn2}
 \ee
 \bea
 && \frac{\ddot a_n}{a_n} + \sum_{l\ne n} \frac{\dot a_n \dot
 a_l}{a_n a_l} \nonumber \\
 &&=  \frac{1}{2(D-1)} \Bigl\{ 2 \Lambda_{D}
 +\frac{1}{\prod_{k=1}^D a_k} \sum_{\{p\}} T_{\{p\}} \prod_{\{p\}} a_k
 + \sum_{ \{ l \} } Q_{ \{ l \} }^2
  \frac{ \prod_{j=1}^{l-1} a_j^2 } {\prod_{k=1}^D a_k^2}   \Bigr\}
 \nonumber \\
 && + \frac{1}{2(D-1)} \Bigl\{ \frac{1} {\prod_{k=1}^D a_k }
 \sum_{m=1}^D \sum_{\{p\}} T_{\{p\}} a_m
 \frac{\partial}{\partial a_m} \prod_{\{p\}} a_k
 + \frac{1}{2} \sum_{m=1}^D a_m \frac{\partial}{\partial a_m}
 \sum_{ \{ l \} } Q_{ \{ l \} }^2 \frac{ \prod_{j=1}^{l-1} a_j^2 }
 {\prod_{k=1}^D a_k^2}   \Bigr\}  \nonumber  \\
 &&- \frac{1}{2} \Bigl\{ \frac{1} {\prod_{k=1}^D a_k } \sum_{\{p\}}
 T_{\{p\}} a_n \frac{\partial}{\partial a_n} \prod_{\{p\}} a_k
 + \frac{1}{2} a_n \frac{\partial}{\partial a_n}
 \sum_{ \{ l \} } Q_{ \{ l \} }^2
  \frac{ \prod_{j=1}^{l-1} a_j^2 } {\prod_{k=1}^D a_k^2}
  \Bigr\} ,
  \label{eoman2}
 \eea

For the anisotropic expansion of dimensions, we consider the
global rotational symmetry is broken down to $SO(p) \times
SO(D-p)$ by $(D-p)$-brane gas. What we are interested is the case
with $p=3$. But we consider the general case to see the dependence
of dimensionality on branes and fluxes. For simplicity, we
introduce two kinds of bulk RR potential and brane which are
characterized by $Q_p$, $Q_{D-p}$, $T_p$ and $T_{D-p}$. Taking the
scale factor for $p$ subspace as $a$ and $D-p$ subspace as $b$,
the equations of motion are given by
 \bea
 &&p(p-1) (\frac{\dot a}{a})^2 + (D-p)(D-p-1) (\frac{\dot b}{b})^2 +
 2p(D-p) \frac{\dot a \dot b}{ab}      \nonumber \\
 && = \Lambda_{D}
 + \frac{ T_p}{b^{D-p}}
 +  \frac{ T_{D-p} }{a^p }
 + \frac{1}{2} \frac{Q_p^2}{b^{2(D-p)}}
 + \frac{1}{2} \frac{ Q_{D-p}^2 }{a^{2p}} ,
 \label{eomcons}
 \eea
 \bea
 &&\frac{\ddot a}{a} + (p-1)(\frac{\dot a}{a})^2 + (D-p)
 \frac{\dot a \dot b}{ab}
  = \frac{\Lambda_{D}}{D-1}
  - \frac{D-p-2}{2(D-1)} \frac{T_p }{b^{D-p}}
 + \frac{D-p+1 }{2(D-1)} \frac{T_{D-p}}{a^{p}}    \nonumber  \\
 &&+ \frac{p(D-p+1)+1}{2(D-1)} \frac{Q_{D-p}^2 }{a^{2p}}
   - \frac{(D-p)^2 -1 }{2(D-1)}
  \frac{Q_p^2 }{b^{2(D-p)}}
   , \label{eompa}
 \eea
 \bea
 && \frac{\ddot b}{b} + (D-p-1)(\frac{\dot b}{b})^2 +
 p \frac{\dot a \dot b}{ab}
  = \frac{\Lambda_{D}}{D-1} + \frac{p+1}{2(D-1)}
 \frac{T_p}{b^{D-p}} - \frac{ p-2 }{2(D-1)}
 \frac{T_{D-p}} {a^{p}}    \nonumber  \\
 && + \frac{ (D-p)(p-1) +1 }{2(D-1)} \frac{Q_p^2 }{b^{2(D-p)}}
  - \frac{p^2 -1}{2(D-1)} \frac{ Q_{D-p}^2 }{a^{2p}}
  . \label{eomd-pb}
 \eea

If we define the two volumes of $SO(p)$ and $SO(D-p)$ as
 \be
 \zeta \equiv a^p ,~~~~ \xi \equiv b^{D-p},
 \ee
the second derivative terms in Eqs. (\ref{eompa}) and
(\ref{eomd-pb}) can be written as
 \bea
 \frac{\ddot \zeta}{\zeta} +
 \frac{\dot \zeta \dot \xi}{\zeta \xi}
  = p \Big \{  \frac{\Lambda_{D}}{D-1}
 &+& \frac{D-p+1 }{2(D-1)} \frac{T_{D-p}}{\zeta}
 - \frac{D-p-2}{2(D-1)} \frac{T_p }{\xi}  \nonumber  \\
 &+& \frac{p(D-p-1)+1}{2(D-1)} \frac{Q_{D-p}^2 }{\zeta^2}
   - \frac{(D-p)^2 -1 }{2(D-1)}
  \frac{Q_p^2 }{\xi^2}  \Big \}
   , \label{eomzeta}
 \eea
 \bea
 \frac{\ddot \xi}{\xi} +
 \frac{\dot \zeta \dot \xi}{\zeta \xi}
  = (D-p) \Big \{
  \frac{\Lambda_{D}}{D-1} &+& \frac{p+1}{2(D-1)}
 \frac{T_p}{\xi} - \frac{ p-2 }{2(D-1)}
 \frac{T_{D-p}} {\zeta}    \nonumber  \\
  &+& \frac{ (D-p)(p-1) +1 }{2(D-1)} \frac{Q_p^2 }{\xi^2}
  - \frac{p^2 -1}{2(D-1)} \frac{ Q_{D-p}^2 }{\zeta^2}  \Big \}
  . \label{eomxi}
 \eea
Decoupling the variables by introducing a constant $E$ as in the
previous section, we have
 \bea
 \frac{\ddot \zeta}{\zeta}
  = E
 &+& \frac{p(D-p+1) + (D-p)(p-2) }{2(D-1)}
 \frac{T_{D-p}}{\zeta} \nonumber  \\
 &+& \frac{p^2(D-p-1)+p +(D-p)(p^2 -1)}{2(D-1)} \frac{Q_{D-p}^2 }{\zeta^2}
   , \label{deczeta}
 \eea
 \bea
 \frac{\ddot \xi}{\xi}
  = E +  \frac{D-2p}{D-1} \Lambda_{D}
  &+& \frac{(D-p)(p+1)+p(D-p-2)}{2(D-1)}
 \frac{T_p}{\xi}  \nonumber  \\
  &+& \frac{ (D-p)^2(p-1) +D-p +p((D-p)^2 -1) }{2(D-1)} \frac{Q_p^2 }{\xi^2}
  . \label{decxi}
 \eea
As a consistency check, one can recover Eqs (\ref{zetadd}) and
(\ref{xidd}) by taking $p=3$ and $T_p = Q_{D-p} = Q_p = 0$. Note
that the constant $E$ is not an independent parameter. For given
configuration of branes and RR fields, it is related to
$\Lambda_{D}$, $T_p$'s and $Q_p$'s by putting the solution
obtained from the decoupled equations (\ref{deczeta}) and
(\ref{decxi}) into the original ones (\ref{eomzeta}) and
(\ref{eomxi}).

Since Eqs (\ref{deczeta}) and (\ref{decxi}) are non-linear because
of the $Q^2$ terms, finding a closed form of analytic solution
seems very difficult. So we search the possibilities of the
solutions. We analyze the behavior of the two subvolumes by
considering the effective potential ${\ddot \zeta} = -
\frac{dV_{\rm eff}(\zeta)}{d \zeta}$, ${\ddot \xi} = -
\frac{dV_{\rm eff}(\xi)}{d \xi}$. The effective potentials are
obtained as
 \bea
 V_{\rm eff} (\zeta) &=& - \frac{c_0}{2} \zeta^2 - c_1 \zeta -c_2 \ln \zeta
              ,    \\
 V_{\rm eff} (\xi) &=& - \frac{d_0}{2} \xi^2 - d_1 \xi - d_2 \ln \xi ,
 \eea
where
 \bea
 c_0 &=& E ,          \nonumber  \\
 c_1 &=& \frac{p(D-p+1) + (D-p)(p-2) }{2(D-1)}T_{D-p} ,   \\
 c_2 &=& \frac{p^2(D-p-1)+p +(D-p)(p^2 -1)}{2(D-1)} Q_{D-p}^2
   , \nonumber
 \eea
 \bea
 d_0 &=&  E + \frac{D-2p}{D-1} \Lambda_{D} ,
   \nonumber   \\
 d_1 &=& \frac{(D-p)(p+1)+p(D-p-2)}{2(D-1)} T_p ,   \\
 d_2 &=& \frac{ (D-p)^2(p-1) +D-p +p((D-p)^2 -1) }{2(D-1)} Q_p^2
  . \nonumber
 \eea

Now let us examine the possible combinations of brane gas and
background RR flux. To see the contribution of RR potential to the
evolution of brane universe, consider the simple case when
$T_{D-p}$ and $ Q_p$ are nonzero with $p =3$. This corresponds to
the case that we considered in Sec. II with the inclusion of
4-form RR flux in the unwrapped dimensions. For effective
potential of the observed subvolume $V_{\rm eff} (\zeta)$, we have
 \bea
 c_0 &=& E ,          \nonumber  \\
 c_1 &=& \frac{4D-9}{2(D-1)}T_{D-3} ,   \\
 c_2 &=& 0   .        \nonumber
 \eea
The shape of the effective potential for $E>0$ is given in Fig. 1.
In this case the equation of motion for $\zeta$ (Eq.
(\ref{deczeta})) is the same as Eq. (\ref{zetadd}). So $\zeta$
expands forever as far as $E$ is positive.

\begin{figure}
\includegraphics[angle=270 , width=8cm ]{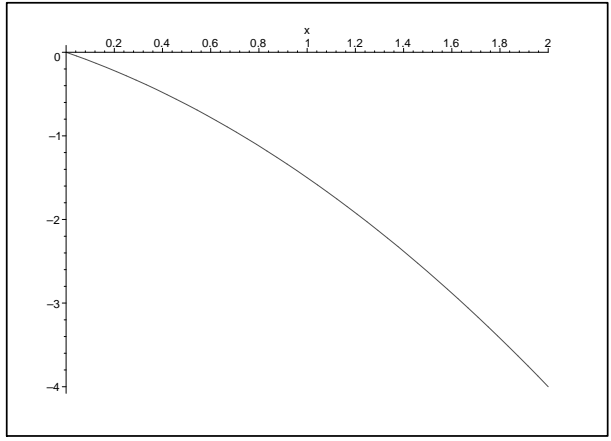} \caption{
Typical shape of effective potential $V_{\rm eff} (\zeta)$ for
unwrapped subvolume for $E>0$. The plot is for $c_0 =1$, $c_1 =
1$, and $c_2 = 0 $. } \label{fig1}
\end{figure}

For effective potential of the internal subvolume $V_{\rm eff}
(\xi)$, we have
 \bea
 d_0 &=& E + \frac{D-6}{D-1} \Lambda_{D} ,
   \nonumber   \\
 d_1 &=& 0 ,   \\
 d_2 &=& \frac{ 5(D-3)^2 + D -6 }{2(D-1)} Q_3^2
  . \nonumber
 \eea
 We need the condition
 $ \Lambda_D = - |\Lambda_D |< 0$ and $E - \frac{D-6}{D-1}
|\Lambda_{D}| < 0$ as in Sec. II to have the confining behavior
for large $\xi$. To have a bouncing potential for small $\xi$, we
need $d_2 >0$ and this condition is automatically is satisfied.
The shape of the effective potential for this case is given in
Fig. 2. The wrapped subvolume $\xi$ will oscillate around the
minimum of the effective potential $\xi_{\rm min} = \sqrt{-d_2/d_0
} = \sqrt{d_2 / |d_0|}$.

\begin{figure}
\includegraphics[angle=270 , width=8cm ]{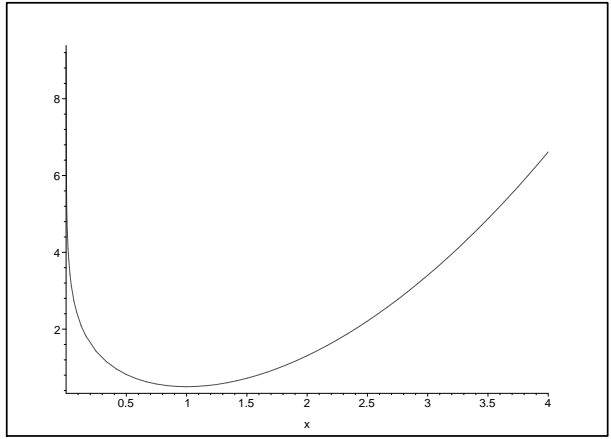} \caption{
Typical shape of effective potential $V_{\rm eff} (\xi)$ for
wrapped subvolume for $ \Lambda_D = - |\Lambda_D |< 0$ and $E -
\frac{D-6}{D-1} |\Lambda_{D}| < 0$. The plot is for $d_0 = -1$,
$d_1 = 0$, and $d_2 = 1 $.} \label{fig2}
\end{figure}

The existence of (3+1)-form RR field induces a logarithmic bounce
on the effective potential of the subvolume perpendicular to it.
This bounce prevents the internal subvolume from collapsing to
zero size. Similarly, the existence of ($D-3+1$)-form RR field in
the internal dimensions can induce nonzero logarithmic bounce for
$\zeta$. However, as far as the condition $E>0$ is satisfied, the
overall behavior of $\zeta$ is essentially the same (see Fig. 3).

\begin{figure}
\includegraphics[angle=270 , width=8cm ]{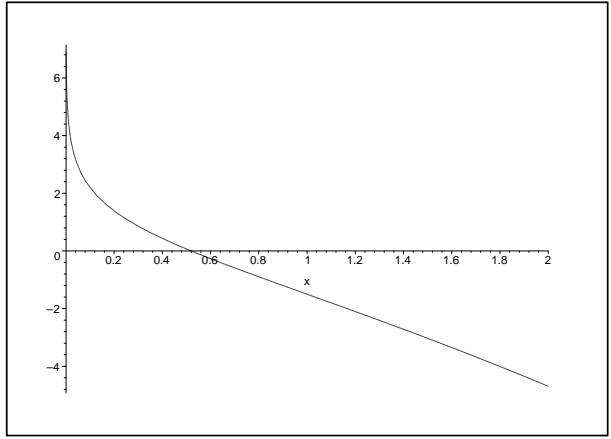} \caption{
Typical shape of effective potential $V_{\rm eff} (\zeta)$ for
unwrapped subvolume when there exists ($D-3+1$)-form RR field in
the internal dimensions. The plot is for $c_0 = 1$, $c_1 = 1$, and
$c_2 = 1 $ } \label{fig3}
\end{figure}

\section{Conclusion}

In the framework of brane gas cosmology, we considered the
asymmetric evolution of the universe. We showed that, provided the
dilaton is stabilized by some mechanism, brane gas and RR
potential can stabilize the internal dimensions. For a combination
of $(D-3)$-brane gas wrapping the extra dimension and 4-form RR
flux in the unwrapped dimensions, it is possible that the wrapped
internal subvolume has an oscillating solution around the minimum
of the effective potential while the unwrapped subvolume expands
monotonically.

One key point of this paper is bulk RR field. The role of this
field is to prevent the internal subvolume from collapsing to
zero. It gives a logarithmic bounce to the effective potential for
small values of internal subvolume. Another is the decoupling of
the observed and internal dimensions by introducing an integration
parameter. With this decoupling, we could treat the problem
analytically in terms of two subvolumes rather than scale factors.

In our analysis we considered only brane gases for the matter part
of the Einstein equation. However, there can be supergravity
particles corresponding to bosonic and fermionic degrees of
freedom. One can ignore the massive modes since these will decay
quickly. For the massless supergravity particles, if we assume
they are homogeneous and isotropic, we can take a perfect fluid
form of energy momentum tensor
 \be
 T^M_N = {\rm diag} (-\rho_{\rm S}, p_{\rm S}, \cdots, p_{\rm S} ).
 \ee
It is well known that this term always tends to drive uniform
expansion. So the main result of our analysis will not be changed
even if we include the massless supergravity particles. To have
more complete stabilization like the one in \cite{patil2}, where
the extra dimensions are stabilized at the self-dual radius, one
can include the effect of other sources. For example, including
the effect of momentum modes or running dilaton may give damped
oscillatory solution in which $\xi$ is driven to the minimum of
the potential.

We would like to point out that the configuration in this work is
just one of the possibilities to achieve the anisotropic
expansion. To make anisotropic expansion between the scale factors
of the observed $SO(3)$ and those of $SO(D-3)$, we considered the
configuration where a gas of positive-tension $(D-3)$-branes wrap
the internal dimension. An alternative way to achieve the same
anisotropic expansion is to use a gas of negative-tension 3-branes
wrapping the observed dimensions. Other combinations of brane gas
and RR flux may give more desirable result.

\begin{acknowledgments}
We would like to thank S. P. Kim for the decoupling of the
equations of motion. We also thank T. Lee and H. W. Lee for useful
discussions.
\end{acknowledgments}

\end{document}